**Title:** Multi-ion, multi-fluid 3-D magnetohydrodynamic simulation of the outer heliosphere

**Short title:** MI-MF simulation of outer heliosphere

**Authors and addresses:**

C. Prested (cprested@bu.edu)
Astronomy Department, Boston University

Astronomy Department
725 Commonwealth Ave.
Boston, MA, 02215

M. Opher
Astronomy Department, Boston University

Astronomy Department
725 Commonwealth Ave.
Boston, MA, 02215

G. Toth
Center for Space Environment Modeling, Department of AOSS, University of Michigan

Department of AOSS
2455 Hayward
Ann Arbor MI, 48109

**Abstract**

Data from the Voyager probes and the Interstellar Boundary Explorer have revealed the importance of pick-up ions (PUI's) in understanding the character and behavior of the outer heliosphere, the region of interaction between the solar wind and the interstellar medium. In the outer heliosphere PUI's carry a large fraction of the thermal pressure, which effects the nature of the termination shock, and they are a dominate component of pressure in the heliosheath. This paper describes the development of a new multi-ion, multi-fluid 3-D magnetohydrodynamic model of the outer heliosphere. This model has the added capability of tracking the individual fluid properties of multiple ion populations. For this initial study two ion populations are modeled: the thermal solar wind ions and PUI's produced in the supersonic solar wind. The model also includes 4 neutral fluids that interact through charge-exchange with the ion fluids. The new multi-ion simulation reproduces the significant heating of PUI's at the termination shock, as inferred from Voyager observations, and provides properties of PUI's in the 3-D heliosheath. The thinning of the heliosheath due to the loss of thermal energy in the heliosheath from PUI and neutral interaction is also quantified. In future work the multi-ion, multi-fluid model will be used to simulate energetic neutral atom (ENA) maps for comparison with the Interstellar Boundary Explorer, particularly at PUI energies of >1 keV.

# 1. Background

The interaction between the solar wind and the interstellar medium (ISM) has been studied via increasingly sophisticated models, where each level of sophistication provides new physical insights. The purpose of this paper is to introduce the first 3-D magnetohydrodynamic (MHD) multi-ion fluid, multi-neutral fluid simulation of the outer heliosphere.

With the increasing availability of computational power, simulations can calculate the interactions between neutrals and ions in a 3-D MHD system. Neutrals and ions interact via charge-exchange, and that communication imprints the interaction of the ISM and solar wind ion fluids onto the neutral fluids. The neutral fluids have distinct properties based on their last interaction with ions.

Papers by Pauls et al. [1995] and Zank et al. [1996] laid down the analytic formulation for the multi-fluid approach in modern simulations, which couples fluids of neutrals and ions via charge-exchange interactions. McNutt et al. [1998] gave alternative formulations of charge-exchange source terms and addressed the accuracy of approximations used in the analytic solution by Pauls and Zank. The McNutt formulation is part of the analytic formulation for the Opher et al. [2009] multi-fluid simulations. This multi-fluid approach implicitly assumes a Maxwellian distribution of neutrals, correspondingly assuming that the mean free path of H-H interactions is smaller than charge-exchange length scales and other physically relevant scales in the simulation.

Isenberg [1986] developed a 1-D three fluid model, consisting of solar wind protons, pick-up ion protons, and electrons, to explain the observed slow down and temperature profile of the supersonic plasma in the outer heliosphere. Wang and Richardson [2001; 2003] continued this work accounting for energy sharing between pick-up ions and the solar wind. Usmanov and Goldstein [2006] extended the three fluid model to 3-D, for distances up to but not including the termination shock.

Models have also provided understanding of the shape and general structures of the heliosphere, including the asymmetric termination shock [e.g. Linde et al. 1998; Ratkiewicz et al. 1998, Pogorelov & Matsuda 1998; Opher et al. 2006] and a wall of enhanced hydrogen density outside of the heliopause [e.g. Baranov et al. 1991; Zank et al. 1996; Fahr et al. 2000]. The evolution in the outer heliosphere of the magnetic sector region, the region where the magnetic polarity flips due to the tilt between the solar magnetic and rotation axes, has been investigated by Czechowski et al. [2010] and Borovikov et al. [2011]. Opher et al. [2011] has recently investigated the magnetic sector region using a multi-fluid approach, increasing the grid resolution in the sector region in the heliosheath.

Neutrals should be treated kinetically, as their mean free paths are ~100 AU, comparable to the size of the heliosheath. Such treatments were done in 2-D by Malama et al. [2006] who treated the interaction between neutrals and ions kinetically using Monte-Carlo simulations, resulting in non-Maxwellian neutral distributions. This treatment changes the density, energy distribution, and flow velocity of the neutral populations. Heerikhuisen et al. [2006] found that the density of the hydrogen wall in their kinetic simulation was twice that of their multi-fluid model, which in turn increased the density of neutrals reaching the inner heliosphere and moved the termination shock inward by 5 AU. However, the general structure of the heliosphere was similar for the Monte-Carlo and the fluid approaches. It was also shown by Alexashov and Izmodenov

[2005] that a 4-fluid treatment of the neutrals can give similar results compared to a kinetic model, where each fluid represents a different physical region in the outer heliosphere. Other works, such Heerikhuisen et al. [2007] and Malama et al. [2006)], coupled a kinetic with a fluid approach. Although this procedure is feasible it is expensive computationally. A hybrid kinetic-MHD simulation, which couples the kinetic treatment with a 4-neutral fluid MHD simulation, performed closer to the fully kinetic simulation than just the fluid treatment [Alouani-Bibi et al. 2011], and future work will consider using a similar treatment.

Fueling the need for even more complex simulations are new observations of the outer heliosphere. In 2004 Voyager 1 crossed the termination shock and into the heliosheath at 94 AU, entering subsonic plasma after a nearly 30 year trip [Stone et al., 2005]. The probe crossed the shock at 34° heliolatitude and within a few degrees of longitude from the heliosphere's nose, the direction of the incoming interstellar medium flow. In 2007 Voyager 2, also crossed the termination shock but at 84 AU at -26° heliolatitude and 33° longitude from the nose [Stone et al. 2008]. The termination shock is asymmetric, 7 AU closer in the south than the north when taking into account solar cycle effects [Richardson et al. 2008]. This asymmetry is likely caused by an inclined interstellar magnetic field, as shown by simulations i.e. Opher et al. 2006; 2007, Pogorelov et al. 2007, Ratkiewicz & Grygorczuk 2008. While Voyager 1's *Plasma Science Experiment* instrument failed before it reached the outer heliosphere, Voyager 2 was able to measure plasma from 10 eV/charge upwards of 6 keV/charge with its plasma instrument and 28 keV/charge to 3.5 MeV/charge with the *Low Energy Charged Particle* (*LECP*) instrument, corresponding to the thermal solar wind and the suprathermal population with a power-law distribution [Krimigis et al. 1977; Decker et al. 2008]. The Voyager plasma instrument is dominated by the more dense thermal plasma and effectively only measures the thermal component in its wide, higher energy channels. Data from the measured plasma components implied the plasma was still supersonic after the shock and the thermal temperature was only $10^5$ K, not $10^6$ K as expected for a shock heated plasma. To make the plasma subsonic, the missing thermal energy must be in a plasma population between the plasma experiment and *LECP* instruments ranges. Within this energy range is where pick-up ions (PUI's) exist. The Voyager data suggests that 80% of the thermal solar wind's ram energy is transferred to the PUI's thermal energy at the termination shock [Richardson et al. 2008].

PUI's with energy of order keV are created upstream of the termination shock. As the solar wind blows outward, it encounters ions recently created from photoionization of neutrals, electron-impact, or charge-exchange between neutrals and ions. The processes of photoionization and electron impact are important at distances less than 10 AU (e.g. Bzowski et al. 2008) and are accounted for in our inner boundary conditions. These new ions are swept up by the solar wind convective electric field and energized to a characteristic speed of twice the solar wind convective velocity, or 1 keV. By the time the solar wind reaches the termination shock, 20-30% of the plasma ions are PUI's (see for example the simulation by Usmanov & Goldstein 2006). Although they represent a fraction of the total ion density, PUI's carry the majority of the thermal pressure just upstream of the termination shock, with a temperature of $10^6$ K compared to the thermal solar wind's temperature of $10^4$ K. In this way, PUI's are an important population for understanding the interaction of the solar wind with the ISM.

Understanding PUI's is also important for interpreting the observations from the Interstellar Boundary Explorer (IBEX) and the *IBEX-Hi* energetic neutral atom (ENA) camera, which has an energy range of 0.7 – 6 keV [McComas et al. 2009; 2011]. ENAs are created when a hot ion in the heliosheath charge exchanges with interstellar neutrals, becoming a hot neutral. Little momentum and energy is exchanged in this process, meaning that the ENAs carry with them the imprint of the plasma from which they are born.

The spectra of the global ENA flux measured by IBEX was ordered with latitude, from ~0.5 - 6 keV, with a distinctly different spectral shape for high latitudes vs. that for low latitudes [Funsten et al. 2009]. IBEX was launched into an earth orbit at the end of 2008, near the end of the solar minimum, and began taking data almost immediately. As it takes the solar wind ~ 9 months to reach the termination shock and a 1 keV ENA at least another 9 months to travel back to IBEX, the first IBEX all-sky maps are a snapshot of the solar-minimum-like heliosheath plasma. During solar minimum, the solar wind velocity is well structured, with fast wind in the high latitudes and slow wind in the equatorial regions. The PUI temperature is dependent on the flow velocity where it was picked up, and it is very likely that the latitudinal ENA spectra ordering observed by IBEX is due to this link between PUI's and the solar wind flow structure. Simulations of the outer heliosphere have the formidable challenge of explaining the magnitude, spectral shape, and geometric structure of the energetic neutral atom (ENA) all-sky maps, and will need to address the impact of PUI's on the IBEX all-sky maps.

Many 3-D simulations of the plasma interaction in the outer heliosphere use a kinetic or multi-fluid treatment of neutrals but only use a single-ion fluid. These simulations capture heating of the total solar wind plasma due to PUI's, but lose the specific properties of the separated PUI and solar wind populations. Prested et al. [2008] described a technique to simulate ENA maps from the properties of plasma and neutrals in MHD simulations, but this technique requires the density, temperature, and flow velocity of each population. It is impossible to self-consistently extract the PUI properties from the single-ion fluid simulation, and this makes it difficult to directly compare the simulation to the IBEX and Voyager observations described above. To address this issue, we make use of the multi-ion capability of the BATS-R-US framework [Glocer et al. 2009; Toth et al. 2012].

The simulation described in this paper is the first 3-D MHD multi-ion, multi-fluid simulation, hence-forth the MI-MF simulation, of the interaction between the ISM and the solar wind. This paper describes a new simulation which includes multi-ion capability, and shows how this new capability immediately improves the simulation's ability to match Voyager 2 observations at the termination shock crossing.

**2. Simulation**

The model builds off of the Opher et al. 2009 single-ion, multi-fluid (SI-MF) simulation, which uses the BATS-R-US framework [Toth et al. 2012]. Full descriptions of the SI-MF and its equations of state can be found in Alouani-Bibi et al. 2011. The cubic grid has a spherical inner boundary of radius 30 AU and a cubical outer boundary at 1500 AU. The grid provides higher resolution near the inner boundary, where resolution is 0.5 AU, and in the nose direction, where resolution is 1.5 AU near the termination shock.

The solar wind's inner boundary parameters are 0.00874 cm$^{-3}$ and 2 x 10$^4$ K, and the PUI's are 0.000945 cm$^{-3}$ and 8.202 x 10$^5$ K. The parameters for the density, velocity, and temperature of the thermal ions and neutrals in the interstellar medium reflect the best observational values. The parameters for the inner boundary (located at 30 AU) were chosen to match those used by Izmodenov et al. [2009]. For the PUI density and temperature we used the 2-D upstream model of PUI's and solar wind by Usmanov and Goldstein [2006]. PUI's comprise 10% of the number density at the inner boundary, and the plasma flows with a radial speed of 417 km/s. The total thermal pressure at the inner boundary in the two simulations was approximately the same in the MI-MF and SI-MF models, but the total ion density was 10% higher in the MI-MF simulation, causing the ram pressure to be 10% higher at the inner boundary. The ISM magnetic field is 4.4 µG and orientated with α=20°, β= 60˚, angles defined in Opher et al. 2009. This orientation was chosen as it reproduces the termination shock asymmetry as observed by the Voyager probes. The ISM plasma at the outer boundary of the simulation has an ion density of 0.06 cm$^{-3}$, neutral density of 0.18 cm$^{-3}$, temperature of 6519 K, and a flow speed of 26.4 km/s with respect to the sun, which is the same for both the MI-MF and SI-MF models.

The MI-MF simulation has 4 neutral fluids, each with properties related to the region of their last interaction with plasma. Region 1 neutrals are created in the region outside the heliopause where the ISM interacts with the heliosphere, region 2 neutrals are created in the heliosheath, region 3 neutrals are created in the upstream solar wind, and region 4 neutrals belong to the pristine ISM. Using this identification system, one can classify different populations of PUI's, with characteristics based on the plasma flow in the region that they are generated. Region 3 PUI's have a thermal energy of 1 keV, from being picked up in the ~400 km/s fast super-sonic solar wind flow, where region 2 PUI's have cooler thermal energy, 0.2 keV, from the ~100 km/s heliosheath flows. For this first MI-MF simulation, only region 3 PUI's are tracked as a separate ion fluid. The region 2 PUI's will be studied in future work.

When a PUI is newly created, it experiences acceleration from the convective electric field of the plasma flowing past it. This acceleration continues until PUI is co-moving with the ambient plasma and convective electric field is no longer present, typically within a few PUI gyro-orbits. Therefore, we assume that the two ion fluids are co-moving and frozen-in with the magnetic field.

On relevant length and time scales, appropriate for a non-thermal PUI population, the PUI's have an energy distribution shape, upstream of the termination shock, similar to the Vasylinus-Siscoe distribution [Vaslyinus & Siscoe 1976], but the energy distribution of PUI's in the heliosheath has not been observed. Particle-in cell simulations by Wu et al. [2010] have shown that sum of the thermal solar wind and non-thermal PUI distributions downstream of the termination shock can be approximated with a 2-Maxwellian distribution. It is important to note that this 2-Maxwellian distribution neglects the suprathermal tail population that has a characteristic power-law distribution. It is not known where the suprathermal tail begins in the heliosheath, but the tail becomes the dominate population for energies >7 keV at a few AU [Fisk & Gloeckler 2006]. The evolution of this transition with radial distance is not well defined, but may become important in the highest IBEX energy channels. We assume that modeling the plasma as two Maxwellians, with properties of the thermal solar wind and PUI's, is a valid first approximation. The suprathermal tail is not included currently in the MI-MF, but future work could incorporate non-thermal plasma populations into the simulation.

The total ion distribution of the MI-MF simulation with two ion fluids is characterized by the summation of the two Maxwellian distributions. An example of the new total plasma distribution and how it compares to the SI-MF simulation in the reference frame of an earth based observer is shown in Figure 1, for plasma with properties taken from the simulation immediately downstream of the termination shock, near the nose. The solar wind and PUI parameters are 7.5 x $10^5$ K, 0.0039 cm$^{-3}$ and 7.6 x $10^6$ K, 0.0013 cm$^{-3}$, respectively, and the plasma flow speed is 110 km/s. In the SI-MF case, the simulation assumes a Maxwellian plasma distribution with the total plasma properties, or the average plasma properties of the solar wind plus PUI's. Figure 1 also shows the effective plasma distribution over two of *IBEX-Lo*'s energy channels and all six of *IBEX-Hi*'s energy channels, represented as stars on the plot. Below 0.1 keV, ENAs produced from the plasma distribution would undergo measureable modulation from the balance between gravitational and radiation pressure, which are important for distances less than 10 AU and are outside the current scope of the MI-MF simulation.

Four neutral fluids are used as in the previous SI-MF simulation. However, because there are two separate ion fluids, additional attention is needed for describing the interaction between ion and neutral fluids. Each ion fluid has distinct properties and produces a different population of neutrals, which are formed during the ion-neutral charge-exchange interaction in a given location. The issue remains how to reconcile the potentially 8 neutral fluids, ideally produced in 4 regions by 2 ion fluids, with the 4 simulated neutral fluids. We allow thermal ions to interact with neutrals, exchanging energy and momentum in addition to creating PUIs in the supersonic solar wind, but we allow only for the loss of PUI's via charge-exchange interactions with neutrals. This approach includes the important loss of thermal energy from the interaction of PUI's and neutrals in the heliosheath [Alexashov & Izmodenov 2005; Malama et al. 2006] while avoiding the creation of additional neutral fluids. Other effects ignored by this approach, such as the creation of PUI's in region 2, are secondary and will be considered in future work.

The following equations describe the dynamics of the ion fluids and the neutral fluids. These equations are similar to those in Alouani-Bibi et al. 2011, with modifications for a second ion fluid. Equations 1-11 apply to region 3, in the supersonic solar wind, distinct in that it is the region in which PUI's are produced. In region 3, the PUI fluid increases in density and energy, and the solar wind fluid loses density and energy as a solar wind ion is lost for every PUI produced. In region 2, the solar wind fluid source functions includes both gain and loss terms, as the solar wind fluid includes the energy and momentum changes from the production of region 2 type PUI's. Outside of the heliopause, in regions 1 and 4, the solar wind subscripts in Equations 12-15 are replaced by ISM subscripts, but the form of the equations is identical.

Below, $\rho$ is mass density, $p$ is thermal pressure, $E = \frac{1}{2}\rho u^2 + \frac{p}{\gamma-1}$ is the total hydrodynamic energy density, $u$ is the flow velocity of the fluid, $U_{th}$ is the characteristic thermal velocity of a particle, $\sigma$ is the charge-exchange cross section as defined by Lindsay and Stebbings [2005], B is the magnetic field, $U_{rel}$ is the relative bulk flow velocity between two fluids, and S stands for the various source terms. Subscripts are SW for solar wind, PUI for pick-up ion fluid, and $H_j$ for the hydrogen neutral fluid produced the $j^{th}$ region: j=1, 2, 3, or 4. The loss of PUI's due to secondary charge-exchange with neutrals is included, but the production of a new ion population from this interaction is not. Therefore there is a small density sink in the total ion fluid and neutral fluids.

The equations for the plasma in the supersonic solar wind (region 3) are:

$$\frac{\partial \rho_{SW}}{\partial t} + \vec{\nabla} \cdot (\rho_{SW}\vec{u}) = S_{\rho,SW} \qquad (1)$$

$$\frac{\partial \rho_{PUI}}{\partial t} + \vec{\nabla} \cdot (\rho_{PUI}\vec{u}) = S_{\rho,PUI} \qquad (2)$$

$$\frac{\partial ((\rho_{SW}+\rho_{PUI})\vec{u})}{\partial t} + \vec{\nabla} \cdot ((\rho_{SW}+\rho_{PUI})\vec{u}\vec{u}) + \vec{\nabla} \cdot \left(\left(2p_{SW} + p_{PUI} + \frac{B^2}{2\mu_0}\right)\overleftrightarrow{I} - \frac{\vec{B}\vec{B}}{\mu_0}\right) = S_M \qquad (3)$$

$$\frac{\partial E_{SW}}{\partial t} + \vec{\nabla} \cdot ((E_{SW}+p_{SW})\vec{u}) + \frac{\rho_{SW}}{\rho_{SW}+\rho_{PUI}}\vec{u} \cdot (-\vec{J}x\vec{B} + \nabla p_{sw}) = S_{E,SW} + \varepsilon S_{E,PUI} \qquad (4)$$

$$\frac{\partial E_{PUI}}{\partial t} + \vec{\nabla} \cdot ((E_{PUI}+p_{PUI})\vec{u}) + \frac{\rho_{PUI}}{\rho_{SW}+\rho_{PUI}}\vec{u} \cdot (-\vec{J}x\vec{B} + \nabla p_{sw}) = (1-\varepsilon)S_{E,PUI} \qquad (5)$$

where the source terms, S, are:

$$S_{\rho,SW} = -q_{ex} = -\sum_{k=1}^{4} \sigma(H_k, U^*_{\rho,H_k,SW})\rho_{SW} n_{H_k} U^*_{\rho,H_k,SW} \qquad (6)$$

$$S_{\rho,PUI} = q_{ex} + q_{ph} - \sum_{k=1}^{4} \sigma(H_k, U^*_{\rho,H_k,PUI})\rho_{PUI} n_{H_k} U^*_{\rho,H_k,PUI} \qquad (7)$$

$$S_M = -\sum_{k=1}^{4} n_{H_k}[\sigma(H_k, U^*_{M,H_k,SW})\rho_{SW} U^*_{M,H_k,SW}(\vec{u}-\vec{u}_H) - \sigma(H_k, U^*_{M,H_k,PUI})\rho_{PUI} U^*_{M,H_k,PUI}\vec{u}] \qquad (8)$$

$$S_{E,SW} = -\sum_{k=1}^{4} \sigma(H_k, U^*_{M,H_k,SW})\rho_{SW} n_{H_k} U^*_{M,H_k,SW} \left[\frac{1}{2}u^2 + \frac{U^*_{\rho,H_k,SW}}{U^*_{M,H_k,SW}} u^2_{th,SW}\right] \qquad (9)$$

$$S_{E,PUI} = \sum_{k=1}^{4} n_{H_k} \left[\sigma(H_k, U^*_{M,H_k,SW})\rho_{SW} U^*_{M,H_k,SW} \left[\frac{1}{2}u^2_{H_k} + \frac{U^*_{\rho,H_k,SW}}{U^*_{M,H_k,SW}} u^2_{th,H_k}\right] - \sigma(H_k, U^*_{M,H_k,PUI})\rho_{PUI} U^*_{M,H_k,PUI} \left[\frac{1}{2}u^2 + \frac{U^*_{\rho,H_k,PUI}}{U^*_{M,H_k,PUI}} u^2_{th,PUI}\right]\right] \qquad (10)$$

where

$$U^*_{M,H_k,SW} = \sqrt{U^2_{rel} + \frac{64}{\pi}U^2_{th,H_k} + U^2_{th,SW}} \qquad (11)$$

$$U^*_{\rho,H_k,SW} = \sqrt{U^2_{rel} + \frac{4}{\pi}U^2_{th,H_k} + U^2_{th,SW}} \qquad (12)$$

$$U^*_{M,H_k,PUI} = \sqrt{U^2_{rel} + \frac{64}{\pi}U^2_{th,H_k} + U^2_{th,PUI}} \qquad (13)$$

$$U^*_{\rho,H_k,PUI} = \sqrt{U^2_{rel} + \frac{4}{\pi}U^2_{th,H_k} + U^2_{th,PUI}} \qquad (14)$$

and the induction equation is:

$$\frac{\partial \vec{B}}{\partial t} - \vec{\nabla} \times (\vec{u} \times \vec{B}) = 0 \qquad (15)$$

The electron pressure is approximated as the equal to the thermal solar wind pressure and $q_{ph}$, the photoionization rate, is approximated to 0, which are appropriate approximations for an inner boundary at 30 AU. The charge exchange rate, $q_{ex}$, is defined in Equation (6). Equations (11)-(14) are derived in McNutt et al, 1998.

The total momentum for the combined plasma fluid is solved for in Equation (3), enforcing that the two ion fluids are co-moving. In the simulation, we have assumed the solar wind and PUI's are distinct fluids, co-moving at speed u. In reality the co-moving plasma should be considered as a mixture of 2 plasmas, particularly at discontinuities and shocks. Also, the MI-MF simulation does not include multiple reflections across the shock, shock surfing, or other small scale shock physics. The energy equations, Equations (4) and (5), are written for hydrodynamic energy density. Note, because the two ion fluids are co-moving we could solve instead for the total energy density including the magnetic density. In solving for the total energy density, we would still have freedom to distribute the total thermal energy between the solar wind and PUI's. This will be addressed in a future work. In Appendix A, we address the jump conditions of the two plasma fluids at the termination shock. The parameter ε is introduced to account for the transfer of heat from PUI's to the solar wind, but in the first iteration of the MI-MF model, ε = 0. The transfer of energy from the PUI's to the solar wind is expected to be a ~5% effect [Malama et al. 2006].

Outside of region 3, the density and energy source terms become:
$$S_{\rho,SW} = 0 \qquad (16)$$

$$S_{\rho,PUI} = -\sum_{k=1}^{4} \sigma(H_k, U^*_{\rho,H_k,PUI}) \rho_{PUI} n_{H_k} U^*_{\rho,H_k,PUI} \qquad (17)$$

$$S_{E,SW} = \sum_{k=1}^{4} \sigma(H_k, U^*_{M,H_k,SW}) \rho_{SW} n_{H_k} U^*_{M,H_k,SW} \left[ \frac{u^2_{H_k} - u^2}{2} + \frac{U^*_{\rho,H_k,SW}}{U^*_{M,H_k,SW}} (u^2_{th,H_k} - u^2_{th,SW}) \right] \qquad (18)$$

$$S_{E,PUI} = -\sum_{k=1}^{4} \sigma(H_k, U^*_{M,H_k,PUI}) \rho_{PUI} n_{H_k} U^*_{M,H_k,PUI} \left[ \frac{1}{2} u^2 + \frac{U^*_{\rho,H_k,PUI}}{U^*_{M,H_k,PUI}} u^2_{th,PUI} \right] \qquad (19)$$

Completing the system of equations the hydrodynamic equations for the neutral fluids are:

$$\frac{\partial \rho_{H_j}}{\partial t} + \vec{\nabla} \cdot (\rho_{H_j} \vec{u}_{H_j}) = S_{\rho,H_j} \qquad (20)$$

$$\frac{\partial (\rho_{H_j} \vec{u}_{H_j})}{\partial t} + \vec{\nabla} \cdot (\rho_{H_j} \vec{u}_{H_j} \vec{u}_{H_j}) + \vec{\nabla} \cdot (p_{H_j} \overleftrightarrow{I}) = S_{M,H_j} \qquad (21)$$

$$\frac{\partial E_{H_j}}{\partial t} + \vec{\nabla} \cdot (E_{H_j} \vec{u}_{H_j}) + \vec{\nabla} \cdot (p_{H_j} \vec{u}_{H_j}) = S_{E,H_j} \qquad (22)$$

For a neutral $H_j$, if the region is also region j, then $\chi = 1$ and $H_j$ are produced and destroyed, else $\chi = 0$ and $H_j$ are only destroyed.

$$S_{\rho H_j} = \chi \left( \sum_{k=1}^{4} \sigma(H_k, U^*_{\rho,H_k,SW}) \rho_{SW} n_{H_k} U^*_{\rho,H_k,SW} \right) -$$
$$n_{H_j} \left( \sigma(H_j, U^*_{\rho,H_j,SW}) \rho_{SW} U^*_{\rho,H_j,SW} + \sigma(H_j, U^*_{\rho,H_j,PUI}) \rho_{PUI} U^*_{\rho,H_j,PUI} \right) \qquad (23)$$

$$S_{M_{H_j}} = \chi(\Sigma_{k=1}^4 \sigma(H_k, U^*_{M,H_k,SW})\rho_{SW} n_{H_k} U^*_{M,H_k,SW} \vec{u}) -$$
$$n_{H_j} \vec{u}_{H_j} \left( \sigma\left(H_j, U^*_{M,H_j,SW}\right) \rho_{SW} U^*_{M,H_j,SW} + \sigma(H_j, U^*_{M,H_j,PUI}) \rho_{PUI} U^*_{M,H_j,PUI} \right) \quad (24)$$

$$S_{E_{H_j},SW} = \chi \left( \Sigma_{k=1}^4 \sigma(H_k, U^*_{M,H_k,SW}) \rho_{SW} n_{H_k} U^*_{M,H_k,SW} \left[ \frac{u^2}{2} + \frac{U^*_{H_k,SW}}{U^*_{M,H_k,SW}} v^2_{th_{SW}} \right] \right) -$$
$$\sigma\left(H_k, U^*_{M,H_j,SW}\right) \rho_{SW} n_{H_j} U^*_{M,H_j,SW} \left[ \frac{u^2_{H_j}}{2} + \frac{U^*_{H_j,SW}}{U^*_{M,H_j,SW}} v^2_{th,H_j} \right] -$$
$$\sigma(H_k, U^*_{M,H_j,PUI}) \rho_{PUI} n_{H_j} U^*_{M,H_j,PUI} \left[ \frac{u^2_{H_j}}{2} + \frac{U^*_{H_j,PUI}}{U^*_{M,H_j,PUI}} v^2_{th,H_j} \right] \quad (25)$$

## 3. Results
*3a) Comparison of SI-MF and MI-MF simulations*

  The comparison of 2-D meridional slices between the SI-MF and MI-MF simulations in Figure 2 shows magnetic field strength and plasma temperatures. The ISM flows from left to right in the figure and the vertical axis is the north-south line. The magnetic field and temperature parameters show the general structure of the heliosphere. In the temperature plots, the termination shock is located at the transition from cooler upstream plasma to hotter downstream plasma. The heliopause is marked by the transition from hot heliosheath plasma to the cooler ISM plasma. The position of the termination shock and heliopause in the Voyager 1 and Voyager 2 directions for the SI-MF and MI-MF simulations are given in Table 1.

  Figure 3 shows the line-cuts of density (3a), total plasma temperature (3b), and plasma flow speed (3c) along the trajectories of Voyager 1 and Voyager 2. The total plasma temperature for the MI-MF is: $T_{plasma} = (n_{SW}T_{SW} + n_{PUI}T_{PUI})/(n_{SW} + n_{PUI})$. The termination shock occurs where the sharp increase in density, increase in temperature, and decrease in flow velocity occurs. The heliopause is most easily seen in the increase in density and decrease in temperature as the plasma transitions from solar wind properties to ISM properties. The line-cut curves are similar away from the termination shock and heliopause with a few notable exceptions. The density is slightly higher in the MI-MF's supersonic plasma, reflecting the 10% density increase at the inner boundary compared to the SI-MF. The temperature curve in Figure 3 falls off more steeply in the MI-MF's heliosheath, implying that there is an increase in thermal pressure loss.

  The distances in Table 1 are the midpoints in the sharp transitions of density and temperature near the TS and HP, and the error is the half width of that transition region. The Voyager 1 transition regions are thicker than at Voyager 2, resulting in higher uncertainty in the distances. The 1-2 AU change in the termination shock location can be attributed to either uncertainty in the simulation's shock position or the 10% increase in the MI-MF plasma density at its inner boundary, which correspondingly increases the ram pressure by 10%. The 7 AU termination shock asymmetry observed by the Voyager probes is preserved within uncertainties. The heliopause position moves significantly inward in the MI-MF simulation, due to an increase loss of thermal pressure in the heliosheath, as discussed further in this section.

  In the SI-MF simulation, the total plasma's thermal velocity, 100 km/s , is used in the calculation of the velocities $U^*\rho$ and $U^*_M$. In the MI-MF simulation, the source terms for the loss of pick-up ions depend instead on the pick-up thermal velocity, 400 km/s. In the heliosheath the plasma flow speed, u, is 100 km/s and the neutral flow and thermal velocities are typically less than 100 km/s. In the MI-MF simulation, the $U^*_{PUI}$ velocity is significantly increased by the large PUI thermal velocity. The thermal speed appears in the plasma density, momentum, and energy source terms (in Equations (6)-(10) and (13)-(15)), which are proportional to $U^*$. The thermal speed also appears a second time in the energy source terms, making the change in energy even more dependent on the thermal speed. In splitting the plasma into 2 populations, the magnitude of the source terms have been changed in the MI-MF simulation, causing more thermal pressure loss in the heliosheath and bringing the heliopause inward.

The heliosheath is dominated by thermal pressure, the majority of which is carried by PUI's. When a PUI charge exchanges with a neutral, a significant parcel of thermal pressure is carried away from the heliosheath by the new neutral. This mechanism is discussed in Malama et al. [2006] for a simulation treating the PUI's as a separate kinetic component from the solar wind fluid. Malama et al. found the termination shock moved 5 AU outward and the heliopause moved 12 AU inward compared to a simulation that treated the plasma as a single ion fluid. These values are similar to what we find in our simulation. The increase in the thermal velocity used to calculate the PUI source terms causes this rate of charge-exchange to increase, increasing the loss of thermal pressure from PUIs in the heliosheath in the MI-MF simulation. The cooling of the plasma in the heliosheath due to this process can be seen in the total plasma temperature curve in Figure 3 and the thermal pressure component in Figure 4.

Figure 4 breaks apart the components of the pressure, ram pressure, thermal pressure, and magnetic pressure in the Voyager 2 direction, and compares the MI-MF and SI-MF simulations. While in the supersonic solar wind, the ram pressure is dominant, in the heliosheath the thermal and ram pressures are comparable. Near the heliopause the magnetic pressure becomes important. The total heliosheath thickness decreased by 18 AU in the Voyager 1 direction and 11 AU in the Voyager 2 direction.

Although the heliopause's are in different locations, the components of plasma flow in the heliosheath behave similarly between the MI-MF simulation and the SI-MF simulation, as shown in Figure 5 for the Voyager 1 and 2 trajectories. Away from the boundaries, the radial, normal, and transverse components follow the SI-MF curves closely in the heliosheath. Therefore the behavior of the flows in the MI-MF heliosheath is essentially the same as the SI-MF and no major changes in the heliosheath flow behavior have been introduced by simulating two ion fluids vs. one ion fluid.

*3b. Capabilities of the MI-MF simulation*

The MI-MF simulation provides additional information about the behavior of the plasma in the outer heliosphere. When Voyager 2 crossed the termination shock, it found the termination shock to be much different than planetary shocks. The sum of solar wind ram energy and solar wind thermal energy, decreased by 80% at the shock. In planetary shocks, this is approximately conserved. The MI-MF simulation reproduced this decrease in solar wind energy, as shown in Figure 6. The termination shock region is shaded blue to draw attention to this steep decrease in energy. Correspondingly, at the shock there is an increase in the PUI thermal energy. This information cannot be extracted from the SI-MF simulation, as the separate properties of the PUI's are not tracked.

The MI-MF simulation provides information about the evolution of the PUI's in the heliosheath with the use of separate ion fluids. Specifically the density and pressure of PUI's are now available, such as in Figure 7. The density of PUI's is 20% of the total ion density close to upstream of termination shock, in agreement with the 2-D upstream multi-ion model by Usmanov and Goldstein [2006]. The termination shock appears as a bright ring of variable thickness in the density plot. This feature is due to the grid, which has a resolution of 1.5 AU towards the nose and 6 AU at higher latitudes. The thermal pressure of PUI's is a large fraction of the total pressure, the sum of thermal, magnetic, and kinetic pressures, in the heliosheath. This shows quantitatively what was implicitly known, that the PUI's are the dominant population in the heliosheath. With this new information, simulated ENA maps can be generated, in the manner described in Prested et al. [2008; 2010] and the intensity of ENAs produced from PUI's can be quantified throughout the sky. These multi-ion maps are expected to be significantly different from simulated maps generated from the single-ion model, as the self-consistent plasma distribution is radically different.

## 4. Discussion and Conclusions

The need for this multi-ion simulation comes from new observations by the Voyager probes and the IBEX satellite. The all-sky maps generated by IBEX are dominated by the PUI population and energies above a keV. To provide the best comparison, global MHD simulations need to provide information about PUI's throughout the 3-D heliosheath. For the first time, this capability is available, and in future work a direct comparison with the IBEX maps will be made.

This new model has some over simplifications which can be improved upon in the future. PUI's created in region 2 (in the heliosheath) are not accounted for in the MI-MF but should contribute only a small amount of thermal pressure in the heliosheath. Malama et al. [2006] found that region 2 PUI's comprised ~ 1% of the total PUI density in the majority of the heliosheath, except for within a few AU of the heliopause where the ion density ramps up and the charge-exchange rate increases. The flow speeds of the heliosheath are much smaller than in the supersonic solar wind, <125 km/s compared to 400 km/s and therefore the region 2 PUI's are a cooler population than the region 3 PUI's (the one originated in the supersonic solar wind). The thermal pressure of region 2 PUI's is <<1% of the region 3 PUI's and can be neglected for this comparison between SI-MF and MI-MF. Including region 2 PUI's in would push the heliopause outward, somewhat decreasing the heliosheath thinning quantified in this paper, but this effect should be smaller than the 1-2 AU grid resolution near the nose's heliopause. Region 2 PUIs should also contribute to ENAs at energies of 0.1 -0.5 keV, an energy range between the thermal solar wind ions and region 3 PUI's. This will be done in a future work.

The first 3-D multi-ion, multi-fluid MHD simulation of the outer heliosphere has been described and discussed. The termination shock asymmetry is preserved in comparison with the SI-MF simulation. However, the location of the heliopause boundary is significantly dependent on the interaction of ions with neutrals in the heliosheath. In splitting the plasma fluid into two components, the magnitude of the source terms has changed, producing a more physically realistic loss of thermal pressure in the heliosheath due to PUI-neutral charge-exchange. In the MI-MF simulation the heliosheath is 18 AU thinner in the Voyager 1 direction and 11 AU thinner in the Voyager 2 direction compared to the SI-MF model. It should be noted that there is significant uncertainty in the heliopause position, 9-12 AU, as the transition region is wide, and the addition of region 2 PUI's is expected to somewhat mitigate this thinning. Despite the change in heliopause position, in comparison to the previous single-ion, multi-fluid simulation by Opher et al., the flows in the new multi-ion simulation flows behave similarly.

The emphasis in this analysis is on the MI-MF simulation's sensitivity to interaction between PUI's and neutrals in the heliosheath. The PUI-neutral interactions in the heliosheath affect more than just the heliopause location. They also affect the upstream plasma, as the flux of neutrals into the inner heliosphere is dependent on ion-neutral charge-exchange in the heliosheath, and they impact the interaction between the ISM and heliosphere, as the position of the heliopause is changed. In future work the interaction between PUI's and neutrals in the heliosheath will be further explored into the MI-MF simulation, and the next generation MI-MF will have the additional capability of ENAs generated directly from the simulation.

*Acknowledgements*
C. P. and M.O. would like to thanks the staff of the NASA Supercomputer Division at Ames and the Pleiades award SMD-10-1600 that allowed the simulations to be performed. The authors also acknowledge the International Space Science Institute (ISSI) in Bern.

**Appendix A**

In the simulation, we have assumed the solar wind and PUI's are distinct fluids, co-moving at speed u. In reality the co-moving plasma should be considered as a mixture of 2 plasmas, particularly at discontinuities and shocks. In Appendix A, we address the treatment of the two plasma fluids at the termination shock.

For a realistic treatment of a mixture of plasma, ignoring energy densities of the magnetic field and electrons and in a planar geometry, the conservation laws of mass, momentum and energy across a steady-state discontinuity imply the following jump conditions:

$$(\rho_{SW} + \rho_{PUI})u = constant \tag{A.1}$$

$$(\rho_{SW} + \rho_{PUI})u^2 + p_{SW} + p_{PUI} = constant \tag{A.2}$$

$$\frac{u^2}{2} + \frac{\gamma}{\gamma-1}\frac{p_{SW} + p_{PUI}}{\rho_{SW} + \rho_{PUI}} = constant \tag{A.3}$$

where $\rho_{SW+PUI}$ is the density of the mixture, and $p_{SW}$ and $p_{PUI}$ are the pressures of the thermal plasma and the PUI respectively. The pressure $p_{SW}$ includes the pressures of thermal ions and electrons.

Equations (A.1)-(A.3) are valid both at smooth parts of plasma flows and at discontinuities. There are additional equations such as the heat transfer equation for the electron temperature that we ignore here. The distribution function of the PUI's should be evolved in a kinetic fashion and diffusion should be taken into account, depending on the level of turbulence in the solar wind (see Malama et al. 2006 for a detailed discussion). In the MI-MF simulation, we do not currently include small scale physics, such as turbulence and other kinetic processes. However the effect of this small scale physics is important for understanding how the solar wind and PUI's transfer through the termination shock.

It is believed that the shock thickness is less than the gyro-radius of a typical upstream pickup proton, at least by a factor of 10 [Fahr & Chalov 2008], and the mean free path of the ions is much larger than the gyro radius. Therefore it is appropriate to treat the termination shock as a discontinuity. In this case the magnetic moment of a PUI after interaction with a perpendicular or quasi-perpendicular shock is the same as it was before the interaction (Toptygin 1980, Terasawa 1979). This qualification is also the condition for weak scattering at the termination shock.

The system of Equations (A.1)-(A.3) is not closed and an additional relation connecting upstream with unknown downstream is required (i.e. Fahr & Chalov 2008). Fahr and Chalov [2008] derive their additional equation based on different scenarios of PUI behavior at the termination shock. For perpendicular shock they considered a case of a) strong scattering, when the velocity distribution is isotropic just downstream the shock and b) weak scattering, when the velocity distribution becomes isotropic only after passage over some distance downstream the shock.

For case a) they obtained:

$$p_{PUI,2} = s^2 p_{PUI,1} \tag{A.4}$$

where s= $\rho_2/\rho_1$ and the index 2 indicates downstream of the shock while 1 indicates upstream of the shock.

For case b) they obtained:

$$p_{PUI,2} = \frac{s(2s+1)}{3} p_{PUI,1} \tag{A.5}$$

Equation (A.4) or (A.5) is the additional equation used to close the system, depending on the type of scattering at the shock. Malama et al. (2006) used a weak scattering limit in which the downstream distribution of PUI is related to the upstream distribution by $f_{PUI,2} = \sqrt{1/s} f_{PUI,1}$.

In our case we solve for the conservation of energy of each fluids, solar wind and PUI separately (see Equations 4, 5, 14 and 15) instead of solving the energy equation for the plasma mixture as in Equation 3. While enforcing the co-moving fluids condition, an additional equation is not needed to describe how the PUI's go through the termination shock. For each fluid the Rankine-Hugoniot conditions are solved. This approach is not entire correct physically since the fluids exchange energy during the shock crossing, and it is only the total energy of the mixture of the fluids that is conserved. However, the final result using our approach is similar, and the difference quantifiable, as described below.

Instead of Equation (A.3) we solve Equations (A.3$^*$), still neglecting the magnetic field:

$$\frac{u^2}{2} + \frac{\gamma}{\gamma-1} \frac{p_{SW}}{\rho_{SW}} = constant \tag{A.3$^*$}$$

$$\frac{u^2}{2} + \frac{\gamma}{\gamma-1} \frac{p_{PUI}}{\rho_{PUI}} = constant$$

Equation (A.3*) can be combined to relate the downstream pressure $p_{PUI,2}$ to the upstream pressure $p_{PUI,1}$ as,

$$p_{PUI,2} = (p_{SW,2} - s_{SW} p_{SW,1}) \frac{\rho_{PUI,2}}{\rho_{SW,2}} + s_{PUI} p_{PUI,1} \tag{A.6}$$

where $s_{SW} = \rho_{SW,2} / \rho_{SW,1}$ and $s_{PUI} = \rho_{PUI,2} / \rho_{PUI,1}$. For co-moving fluids, $s_{SW} = s_{PUI} = s$, according to mass conservation and neglecting the source terms for the relatively short distance across the shock.

Considering that $\rho_{PUI,2} / \rho_{SW,2} \sim 0.3$, our system of equations gives the following relationship between the downstream to the upstream pressure:

$$p_{PUI,2} = s p_{PUI,1} \tag{A.7}$$

For a weak shock such as observed by the Voyage probes, $s = 2$, we obtain a ratio between upstream and downstream PUI pressure of 2, where as with weak scattering the result is 3.3. Therefore our model heats the PUI's less, 30% in this example, than a model which includes a kinetic treatment of the termination shock. This heating difference should be considered when interpreting our results.

**Figures**

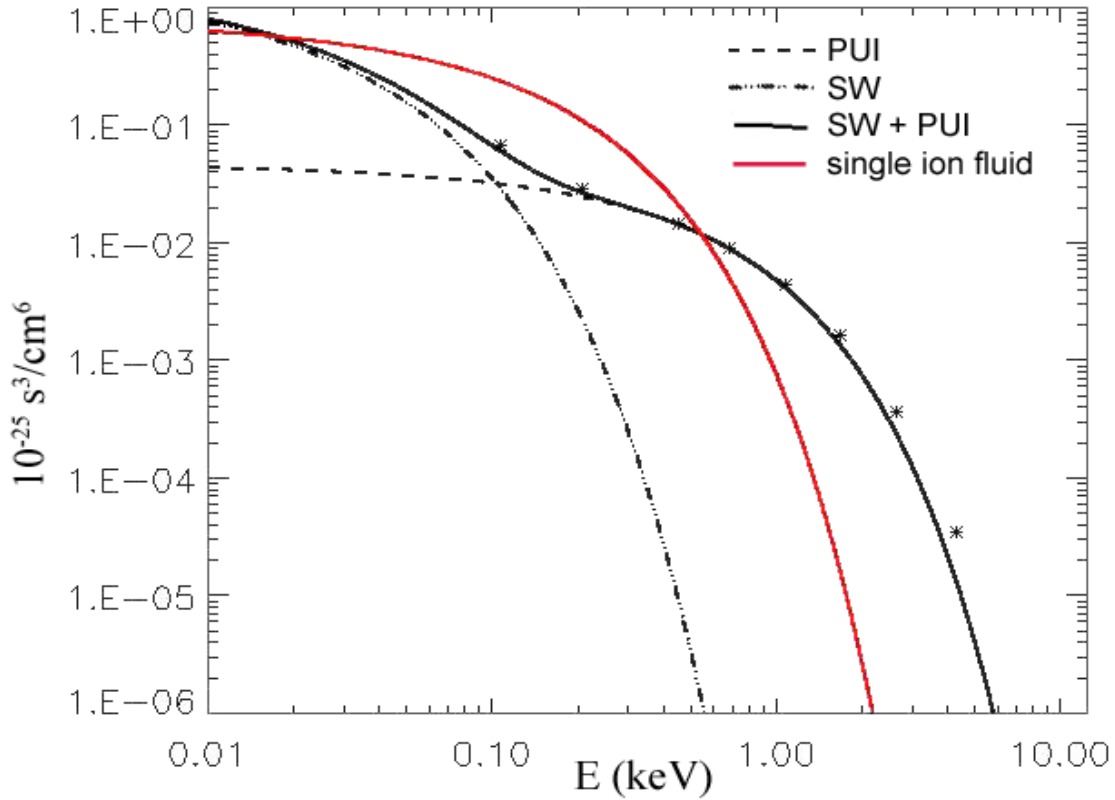

Figure 1: Energy distribution of various plasma components in the frame of the heliospheric observer for values given in the text. The solar wind (dot-dashed line) and PUI (dashed line) are approximated as Maxwellian distributions. The solid black line shows the sum of the two populations, which is significantly different than the solid red line, a single Maxwellian distribution based on the average plasma properties of the combined populations. The stars represent the combined plasma distribution for relevant IBEX energy channels. The energy channels are wider at higher energies, and the effective plasma energy distribution is larger at the central energy than the plasma distribution curve.

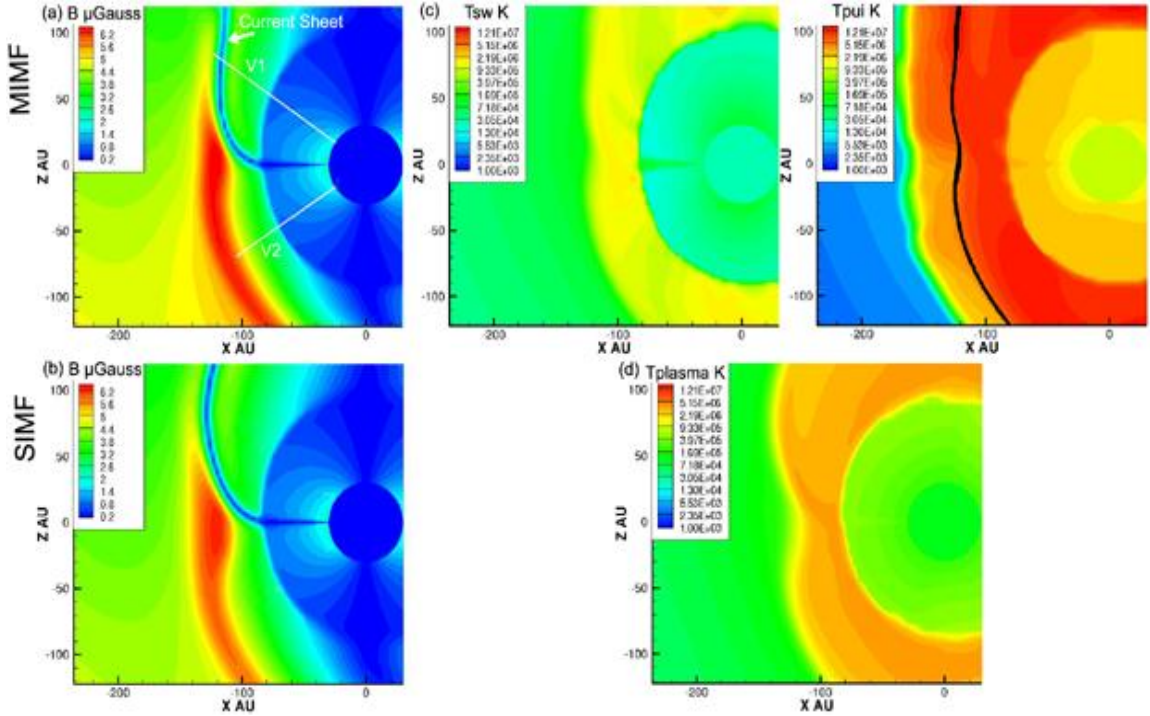

Figure 2: Showing the (a,b) magnetic field intensity for both simulations, (c) solar wind and PUI temperature for the MI-MF simulation, and (d) total plasma temperature for the SI-MF simulation. The current sheet is marked in the top magnetic field plot and the ISM flow comes from the left. Also shown are the projections of the Voyager 1 and Voyager 2 trajectories. In the (c) PUI temperature plot, the location of the heliopause is shown in black. The extension of the temperature of the PUI's is an artifact of the simulation near the heliopause. However, beyond the heliopause, the density of PUI's is very low, $<10^{-5}$ cm$^{-3}$, and the thermal pressure from PUI's is negligible.

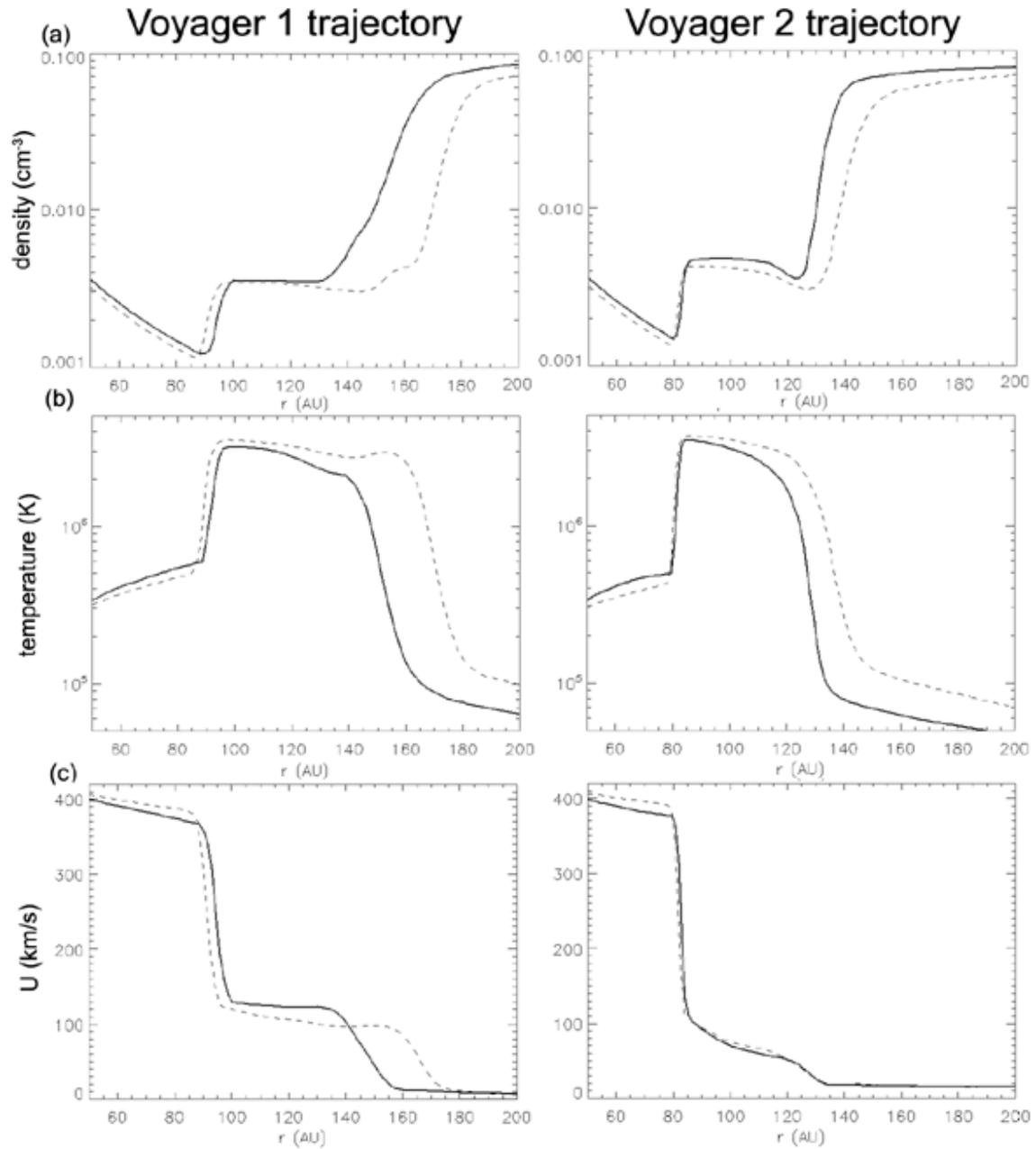

Figure 3: Comparison of (a) density, (b) total plasma temperature, and (c) flow speed along the Voyager 1 and Voyager 2 trajectories. The solid line is the multi-ion simulation and dashed line is the single-ion simulation. The density and temperature plots have a log scale on the y-axis while the flow speed plot is on a linear scale.

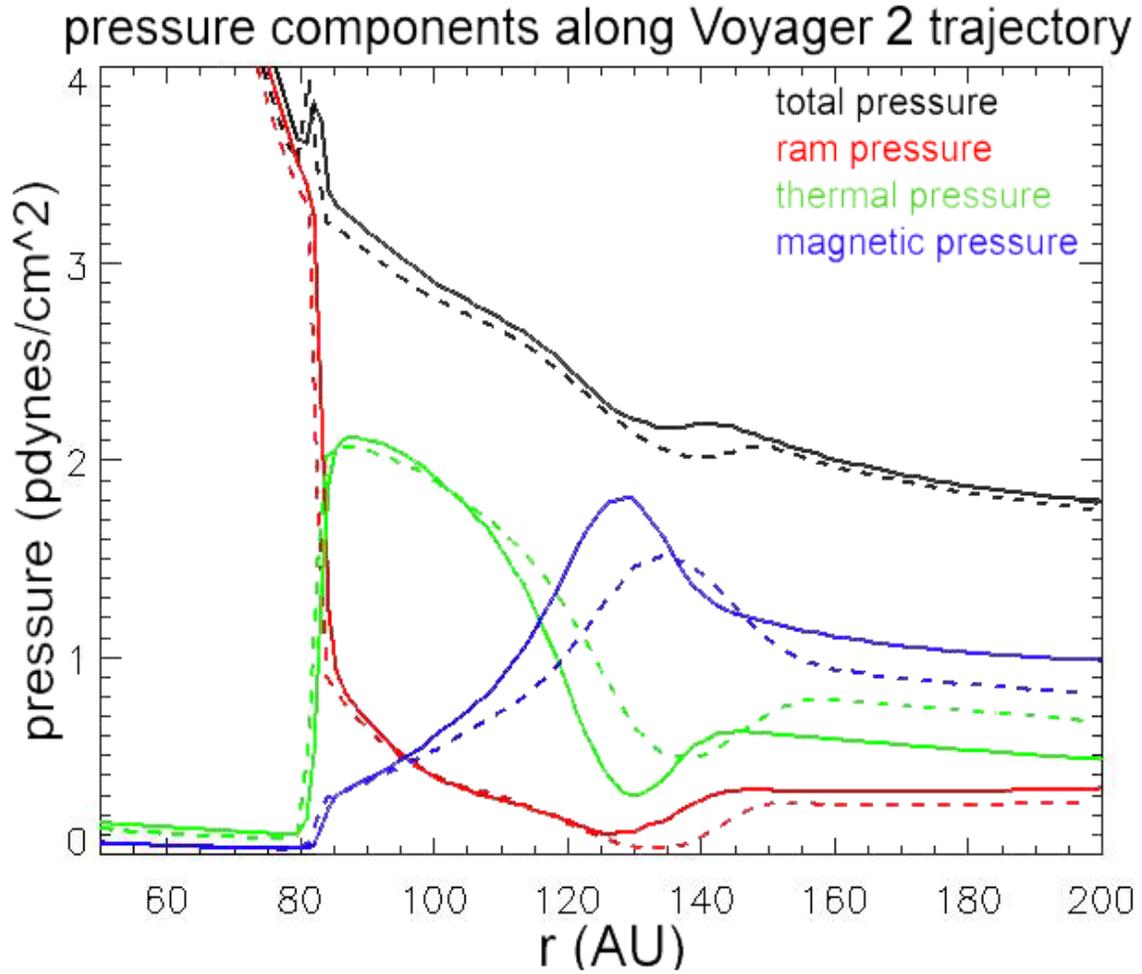

Figure 4: Components of pressure in the heliosheath along the Voyager 2 trajectory. The black line represents the sum of the plasma thermal pressure (green), ram pressure (red), and magnetic pressure (blue) for the MI-MF (solid) and SI-MF (dashed) models.

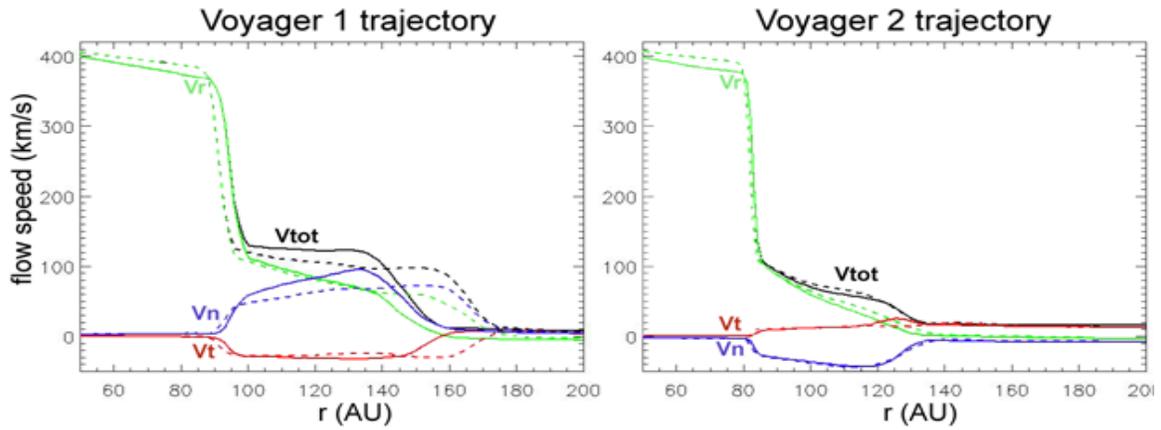

Figure 5: Components of plasma flow velocity along the trajectories of Voyager 1 and 2. Vn, Vr, and Vt are defined in the text. The solid line is MI-MF simulation and dashed line is SI-MF simulation. The flows are very similar in the two simulations.

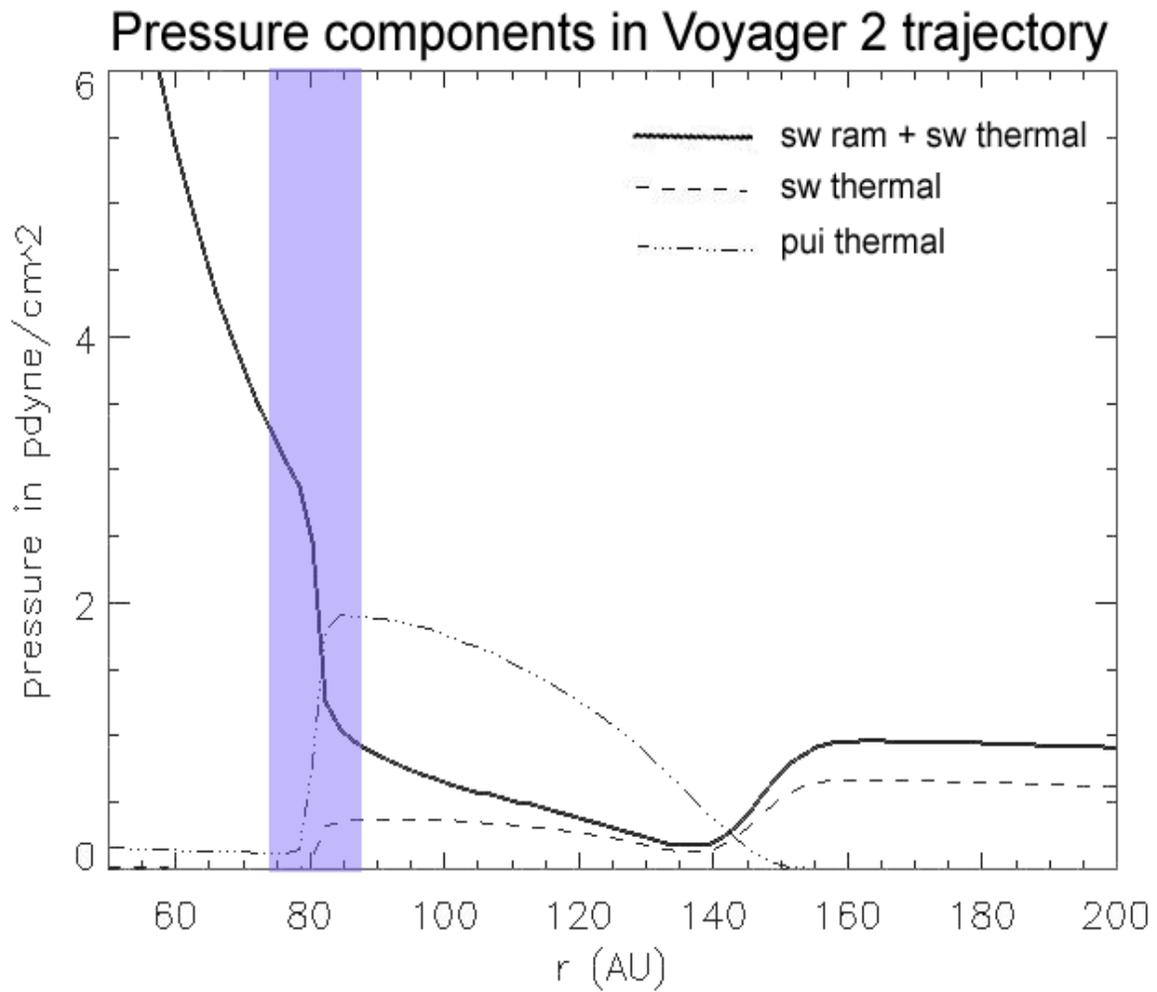

Figure 6: Figure showing the decrease in the total solar wind pressure and the jump in PUI thermal pressure at the termination shock (shaded blue) in the Voyager 2 direction. The simulation replicates the 80% decrease in the thermal solar wind energy per proton (ram plus thermal pressure) as observed by Voyager 2. The decrease in energy/solar wind proton corresponds to an increase in PUI thermal energy, as inferred by Voyager 2 and replicated in the simulation.

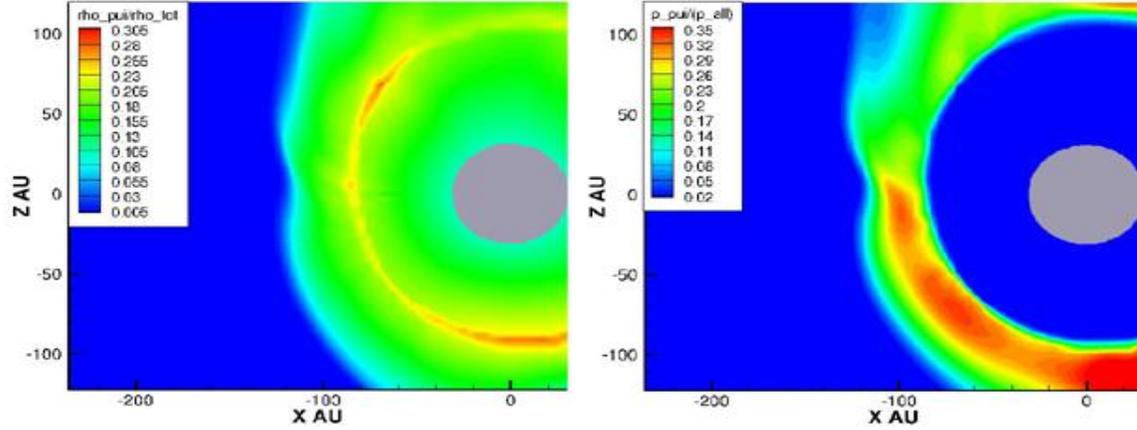

Figure 7: The fractional density of PUI's (left) and the fractional pressure of PUI's (right) in a meridional cut. The fractional density is the PUI density over the total plasma ion density. The fractional pressure is the ratio of the thermal PUI pressure to the sum of the total thermal, magnetic, and ram pressures. In the left panel, the variability in the termination shock is due to the simulation grid spacing, which varies in that region between 1.5-6 AU. The inner boundary region, 30 AU, is shaded gray in the pressure image.

**Tables**

Table 1: Comparison of heliosheaths in single-ion fluid and multi-ion fluid models

| Model | Voyager 1 termination shock | Voyager 1 heliopause | Voyager 2 termination shock | Voyager 2 heliopause |
|---|---|---|---|---|
| SI-MF | 91 ± 4 AU | 166 ± 9 AU | 81 ± 3 AU | 139 ± 9 AU |
| MI-MF | 93 ± 5 AU | 150 ± 12 AU | 82 ± 3 AU | 129 ± 9 AU |